# Searle's "Dualism Revisited"

Henry P. Stapp

*Theoretical Physics Group*
*Lawrence Berkeley National Laboratory*
*University of California*
*Berkeley, California 94705*

**Abstract:** *A recent article in which John Searle claims to refute dualism is examined from a scientific perspective.*

## 1. Introduction.

John Searle begins his recent article "Dualism Revisited" by stating his belief that the philosophical problem of consciousness has a scientific solution. He then claims to refute dualism. It is therefore appropriate to examine his arguments against dualism from a scientific perspective.

Scientific physical theories contain two kinds of descriptions:

(1) Descriptions of our *empirical* findings, expressed in an every-day language that allows us communicate to each other our sensory experiences pertaining to what we have done and what we have learned; and

(2) Descriptions of a *theoretical* model, expressed in a mathematical language that allows us to communicate to each other certain ideas that exist in our mathematical imaginations, and that are believed to represent, within our streams of consciousness, certain aspects of reality that we deem to exist independently of their being perceived by any human observer.

These two parts of our scientific description correspond to the two aspects of our general contemporary dualistic understanding of the total reality in which we are imbedded, namely the empirical-mental aspect and the theoretical-physical aspect. The duality question is whether this general dualistic understanding of ourselves should be regarded as false in some important philosophical or scientific sense.

Searle first contrasts his own anti-dualist approach, which he calls "biological naturalism", to the two forms of materialism, namely reductionism and eliminativism, of which he says:

> I think in the last couple of decades the weaknesses of reductionism and eliminativism have become apparent to most workers in the field. However, an odd thing has happened: dualism has gradually come to seem intellectually respectable again. One of the main aims of this article is to show the incoherence of dualism. Both materialism and dualism are false, …(Searle, 2007)

Although Searle speaks of a scientific solution, some of his presumptions are not in line with contemporary science. For example, he asserts that "All conscious states, without exception, are caused by neurobiological processes in the brain. We now have an overwhelming amount of evidence for this, …"

The empirical evidence from neuroscience, and the other sciences, is far from entailing that all conscious states are caused 100% by neurobiological processes in the brain. The empirical evidence, which is in the form of correlations, does not absolutely rule out the possibility that a person's current mental state could be influenced in some way direct way by his immediately prior mental state; influenced in way that is not completely explained causally by neurobiological processes in the brain. In the context of an argument against a naturalistic science-based dualistic understanding of reality, it is both scientifically inaccurate and philosophically 'question begging' to accept as a *science-based premise* this strong assumption that "All conscious states, without exception, are caused by neurobiological processes in the brain." Searle also assumes, again without data-based justification, that "There is nothing to the causal power of consciousness that cannot be explained by the causal power of the neuronal base."

These two causality conditions can rationally be taken as ideology-based desiderata of some hoped-for theory, but not as empirically validated premises of an argument against dualism, for they are not empirically validated. Indeed, the question of the relationship between states of consciousness and states of the brain is coming under increasingly intense scrutiny by neuroscientists. For example, recent studies by neuroscientists of mind-brain relationships show strong correlations between a person's consciously felt and reported intent---effectively controlled by laboratory procedures specified in psychologically described ways, and communicated to the subjects in psychologically meaningful terms---and subsequent neurobiological processes in their brain. (Ochsner, 2002: Wager, 2008) On the face of it, these experiments suggest that our consciously felt intentions might possibly influence the activities of our brains: that our psychologically described experiential aspects, far from being the epiphenomenal by-product of physically described brain activity, as required by nineteenth century classical physical theory, could *themselves* causally affect the course of brain events.

Of course, *classical* physics does demand that all brain activity be fully determined by prior physically described activities alone. But the dictates of classical physics cannot be relied upon to reveal the whole truth in the case of brain activities. This is because brain dynamics depends critically upon such things as flows of ions into nerve terminals, and the dynamical properties of ions are not correctly specified by classical physics. Absent the presumption of the applicability of classical physics to the ion dynamics in the brain, one cannot conclude from the empirical data either that every brain action is wholly caused by prior brain action alone, or that every conscious thought is completely caused by brain activity alone; the empirical data is unable to reveal what is causing what.

A scientifically more secure premise, in the present state of neuroscience, might be that "All conscious states are caused in part by ongoing neurological processes in the brain and in part by the ongoing field of consciousness", without specifying in advance how large the first part is: it might be the 'whole', but that is something that scientists do not yet know for sure. Yet Searle's argument against dualism dissolves if one replaces his strong philosophical premise by this empirically more secure one.

What is Searle's argument against dualism? In his section 5, entitled "What is wrong with dualism?" he notes that "I have already said that consciousness is not ontologically reducible to brain processes. Isn't that already a kind of dualism? Isn't the irreducibility of consciousness all that dualism amounts to?"

He continues: "It is important in answering this question to remind ourselves that I said that consciousness was causally reducible, but not ontologically reducible, to neuronal processes."

This causal reducibility means *fully* causally reducible, not the mere partial causal reducibility mentioned above, which is the most that the contemporary empirical evidence might entail. If we have only that partial reducibility, then we are back to something very close to Descartes' dualism, with the *admittedly ontologically distinct* mental and physical aspects of the human person interacting with each other within the person's brain. Without his strong, but scientifically unsubstantiated, premise Searle would, because of his explicitly dualistic ontology, seem to end up with a form of dualism. How does he counter that apparent conclusion?

Searle says:

> The real objection to dualism is that we cannot give a coherent account of reality on dualist assumptions. We cannot give an account of reality which makes a part of the real world---our conscious states---cohere with our account of the rest of the real world. Dualism postulates two distinct domains, but on this postulation it becomes impossible to explain the relationship between the two domains. This incoherence has a number of consequences. Perhaps most famously, it becomes difficult, if not impossible, to explain how brain processes in one ontological domain could cause consciousness in another ontological domain. Right now I want to focus on another absurd consequence that I mentioned earlier: Epiphenomenalism. If consciousness has the features of qualitativeness, subjectivity, unity, and intentionality, but is not part of the material or physical world, then how on earth could it possibly function in the physical world? ….[Yet] I decide to raise my arm, … and then the arm goes up. There isn't any doubt that my conscious intention causes my arm to go up.

This conclusion, that a person's conscious intent can cause an intended bodily action to occur, is in line with what was suggested by the neuroscience experiments described earlier. It is also in line with the intuitive understanding of the mind-body connection that is the basis of our entire lives. Searle is undoubtedly justified in concluding that this putative mind-body connection is real, and that an adequate neurobiological theory needs to explain it in a rationally coherent science-based way.

How, then, can we give a rationally coherent, naturalistic, science-based explanation of the capacity of our conscious intentions, which belong to the ontological realm of conscious experiences, to affect brain activities, which belong to the ontological realm of Material/Physical objects and fields?

To answer this question, Searle proceeds by listing the disparate properties of Consciousness and the Material World, respectively, on the left and right sides of a table, and asserts : "But we know that my conscious intention-in-action does cause the movement of my arm.  So what is the solution of the puzzle? I think the solution is obvious: Move the ontologically subjective features on the left hand side over to the right hand side. … But we will find it embarrassing to say subjectivity, etc, are "physical" or "material"….So, let us get rid of the terminology….and just say that qualitativeness, subjectivity are parts of the real world just like everything else."

But dualism itself says that the parts of our human natures that are described in subjective experiential/empirical terms and the parts that are described in objective mathematical/physical terms are both aspects of the full reality: our mental aspects and our physical aspects are both parts of the total reality. Duality says that reality contains these two ontologically different kinds of things, and that they interact in human brains. But this duality approach is straight-forward: It does not try to gloss over the distinction between the two aspects of reality by moving one over to the other, and eliminating the terminology that characterize their differences.

Searle has long insisted that these two aspects are ontologically different. This asserted ontological difference has been the basis of his philosophical position: experiential aspects cannot be reduced to physical aspects because the two aspects have different ontologies. Is this blurring of an essential ontological distinction really a philosophically respectable escape from what appears to be the dualism inherent in Searle's dual-ontology framework?

Searle's only actual argument for this "solution" is that he does not see any other way out. He cannot see a naturalistic, science-based, rationally coherent way to understand the 'obvious' causal power of our thoughts to move our bodies, without, in his own words, returning to the old identity theory:

> "But doesn't that leave us open to the objection that this is just the old identity in disguise?  Aren't we just saying that conscious states are neurobiological states of the brain. Well, in one way it seems to me

> that so stated the identity theory is absolutely right and could hardly be wrong."

But he goes on to say that most identity theorists---that he knows---"wanted to say that consciousness is *nothing but* neurobiological states of the brain described in third-person terms." Searle's claim is that he is asserting that consciousness is *more* than just that!

But if a state of conscious is indeed, as he claims, a high-level third-person neurobiological state of the brain *plus an experiential add-on*, and all causation is carried by the third-person aspects, then the experiential add-on is playing no causal role. But that "solution" just brings us back to the old puzzles. Why does the experiential add-on exist at all if it has no causal power? How can it evolve in a naturalistic way if has no physical consequences? Does not this approach just return us to the absurd epiphenomenalism that Searle has rejected so unequivocally?

This failure of Searle's argument to satisfactorily resolve the old dilemmas stemmed directly from his inability to see any other way to avoid epiphenomenalism: from his inability to see any other way to account for the capacity of our conscious intentions to move our bodies; from his inability to see any other way to account for the causal efficacy of our conscious thoughts in the physically described world. This narrowness of his vision stemmed in turn from the postulated causal determinateness of the physically described world that is enshrined in his premises. But according to contemporary basic physical theory, namely quantum theory, the brain is not actually deterministic in that classically conceived way.

Brain dynamics depends crucially upon the motions of calcium ions into nerve terminals, in connection with the release of neurotransmitter molecules. These ions are so small that the effects of the Heisenberg uncertainty principle must be taken into account. Calculations show that these quantum effects at the level of the nerve terminals must be large. This uncertainty/indeterminateness that enters unavoidably at the nerve-terminal level percolates up to the macro-level via the so-called "butterfly" effect, and this leads to *macroscopic* indeterminateness.

Orthodox quantum mechanics deals with this macroscopic indeterminateness in a very specific way. The macroscopic indefiniteness generated by the microphysical laws is brought into accord with the definiteness of our human experiences by incorporating into the physical dynamics certain very specific causal effects of our own thought-provoked actions.

This radically enlarged and dynamically active role of human beings in the quantum mechanical conception of nature was often strongly emphasized by the founders of quantum mechanics. Thus the closing words of Bohr's first book assert, directly in connection with the problems of life and consciousness with which we are concerned here:

> That a physicist touches upon such questions may perhaps be excused on the ground that the new situation in physics has so forcibly reminded us of the old truth that we are both onlookers and actors in the great drama of existence. (Bohr, 1934, p. 119)

Bohr repeats this many times in his writings, and Heisenberg's penultimate sentences in his chapter "The Copenhagen Interpretation" of his book *Physics and Philosophy*, are:

> Our scientific work in physics consists in asking questions about nature in the language we possess and trying to get an answer from experiment by the means that are at our disposal. In this way quantum theory reminds us, as Bohr has put it, of the old wisdom that when searching for harmony in life one must never forget that we are ourselves both players and spectators. (Heisenberg, 1958, p. 58)

But what are Bohr and Heisenberg both emphasizing in calling human beings not merely spectators, but also actors?

The logic and mathematical structures of the orthodox interpretation of quantum mechanics were put into particularly clear form by the great logician and mathematician John von Neumann. (1932/1955). The essential active role of the human agent was formalized by von Neumann as "process 1". Prior to the occurrence of any increment of knowledge in any person's stream of consciousness, an associated specific question must be posed. The image in the quantum mathematics of the action of posing this question is called *process 1* by von Neumann. *This "process 1" is not caused by any physical process described in the theory. Nor is it subject to the famous element of randomness in quantum mechanics. Yet it can have profound effects upon the subsequent course of physically described events.*

*In actual scientific practice* the choice of which question to pose---of which process 1 action to actualize---comes from the *psychologically described realm of the human agent's interests and conscious intentions.* In setting up the empirical conditions, it is we scientists that, on the basis of our scientific interests, choose the questions that we put to nature. In general, within quantum mechanics, it is this capacity of human beings to choose, on the basis of personal interests and goals, questions naturally associated *within the theory* with physically efficacious actions*,* that gives us human beings, *as we are represented within quantum mechanics*, the power to influence our own lives in ways motivated by our own values and interests.

In orthodox quantum mechanics, this freedom to choose certain actions *constitutes an essential part of the process of removing the aspects of quantum macroscopic indeterminateness* that conflict with the determinateness of our actual experiences. These chosen actions *are not determined by any known principle of physical coercion,* but act *inside* the realm of possibilities generated by Heisenberg's uncertainty principle. In the classical approximation that realm of quantum uncertainty collapses to zero, thereby producing the physical determinateness of classical physics. Given this understanding of how quantum mechanics works, it is completely unreasonable, from a contemporary scientific perspective, to impose the deterministic aspect of the *classical approximation* as a premise governing the causal behavior of the ion-driven dynamics of the brain of a conscious human being.

The odd happening reported by Searle, that dualism has gradually come to seem intellectually respectable again, may be due to the growing recognition among scientists that premises based on classical mechanics may not constitute a perfectly adequate foundation for understanding the relationship between ion-driven brain dynamics and conscious observation. Both ion dynamics and the connection between physical description and conscious observation lie in the province of quantum mechanics!

The technical details of the impact of quantum mechanics on our science-based understanding of the connection between consciousness and brain dynamics have been described in growing detail in several publications. (Schwartz, 2005; Stapp, 1993-2009)   These papers give a detailed putative *dynamical explanation* of how our conscious intentional thoughts tend to produce *the consciously intended* physical consequences.

Models of the brain based on the precepts of classical physics have not produced any comparable explanation of the connection between physically described brain activity and our conscious experiences. The reason for this failure of the classical-mechanics-based approach is that classical mechanics is constructed in a way that leaves out our conscious experiences: classical mechanics has no logical place for, or need for, our streams of conscious experiences. Our human experiences enter as alien passive spectators of a causally closed mechanical universe. Quantum mechanics, on the other hand, is fundamentally and explicitly about the structure of our streams of conscious experiences: it is a set of rules that are designed to allow us to calculate expectations pertaining to future experiences on the basis of knowledge gleaned from prior experiences. It *needs* inputs from our streams of consciousness that *act in specified ways* in the physically described world in order to cut the burgeoning possibilities arising from the quantum uncertainty principle back down to the complexes of possible human experience of the kind that lie in the domain of applicability of the theory. The conceptual structure is a rationally coherent unity in which our streams of conscious experiences and our physically described brains are tightly linked by explicitly specified laws. To ignore these scientific developments that are so profoundly pertinent to the issue of the problem of the mind-brain relationship is not scientifically reasonable, particularly in the light of the persistent failures of the classical-physics-based approaches to provide any comparable, rationally coherent, naturalistic explanation of the empirically observable correlations between appearances and measurable physical properties of the brains that host them.

**References.**

Bohr, Niels, 1934. Atomic theory and the description of nature. Cambridge U. P., Cambridge.

Heisenberg, W., 1958. Physics and Philosophy. Harper and Rowe, New York.

Ochsner, K.N., Bunge, S.A., Gross, J. J., and Gabrieli, J.D.E., 2002. Rethinking feelings: An fMRI study of cognitive regulation of emotion. Journal of Cognitive Neuroscience, 14, 1215-1229.

Searle, J.R., 2007. Duality Re-Visited. Journal of Physiology-Paris 101 (2007) 169-178. (Reprinted In ScienceDirect. Doi:10.1016/j.jphysparis.2007.11.003).

Schwartz, J., Stapp, H., and Beauregard, M., 2005. Quantum Theory in Neuroscience and Psychology: A Neurophysical Model of the Mind-Brain Interaction. Philosophical Transactions of the Royal Society B 360 (1485), 1305-1327.

Stapp, H.P. 1993. Mind, Matter, and Quantum Mechanics, 1$^{st}$ edn. Springer, Berlin, Heidelberg, New York.

Stapp, H. P., 1999. Attention, Intention, and Will in Quantum Physics. J. Consciousness Studies, 6, 143-164. Reprinted in The Volitional Brain, Academic Press, New York. 1999.

Stapp, H. P., 2001. Quantum Mechanics and the Role of Mind in Nature, Found. of Physics 31, 1465-1499.

Stapp, H.P., 2005. Quantum Interactive Dualism: An Alternative to Materialism. J. Consciousness Studies, 12, 43-58.

Stapp, H.P., 2006. Quantum Approaches to Consciousness. In: Cambridge Handbook of Consciousness, M. Moscovitch and P. Zelazo (Eds.). Cambridge U.P., Cambridge.

Stapp, H.P., 2007a. Quantum Mechanical Theories of Consciousness. In: Blackwell Companion to Consciousness. M. Velmans and S. Schneider (Eds.).

Stapp, H. P., 2007b. Mindful Universe: Quantum Mechanics and the Participating Observer. Springer, Berlin, Heidelberg, New York.

Stapp, H. P., 2009a. Physicalism Versus Quantum Mechanics, in Mind, Matter, and Quantum Mechanics, 3$^{rd}$ edn, Chapter 13. Springer, Berlin, Heidelberg, New York. (arxiv.org/abs/0803.1625).

Stapp, H. P., 2009b. A Model of the Quantum-Classical and Mind Brain Connections, and the Role of the Quantum Zeno Effect in the Physical Implementation of Conscious Intent, in Mind, Matter, and Quantum

Mechanics, 3$^{rd}$ edn, Chapter 14. Springer, Berlin, Heidelberg, New York. (arxiv.org/abs/0803.1633).

Stapp, H. P., 2009c. Philosophy of Mind and the Problem of Free Will in the Light of Quantum Mechanics. To appear in the book *On Thinking*, Springer, Berlin, Heidelberg, New York. (arxiv.org/abs/0805.0116).

Wager, T., Davidson, M., Hughes, B., Lindquist, M., Ochsner, K., 2008. Prefrontal-Subcortical Pathways Mediating Succesful Emotion Regulation, Neuron, 59, 1037-1050.

Von Neumann, J., 1932/1955. Mathematical Foundations of Quantum Mechanics. Princeton U.P., Princeton.

____________________________________________________________
**This work was supported by the Director, Office of Science, Office of High Energy and Nuclear Physics, of the U.S. Department of Energy under contract DE-AC02-05CH11231**